\documentclass[10pt,conference]{IEEEtran}
\pdfoutput=1
\IEEEoverridecommandlockouts
\usepackage{cite}
\usepackage{amsmath,amssymb,amsfonts}
\usepackage{algorithmic}
\usepackage{graphicx}
\usepackage{textcomp}
\usepackage{xcolor}
\usepackage{ulem}
\usepackage{CJKutf8}
\usepackage{subfigure}
\usepackage{multirow}
\usepackage{multicol} 
\usepackage{arydshln}
\usepackage{booktabs}
\usepackage[top=0.70in,left=0.68in,right=0.68in,bottom=0.95in]{geometry}
\usepackage[colorlinks=true,linkcolor=black,urlcolor=black,citecolor=black]{hyperref}
\newcommand{\CLASSINPUTtoptextmargin}{0.7in}

\def\BibTeX{{\rm B\kern-.05em{\sc i\kern-.025em b}\kern-.08em
    T\kern-.1667em\lower.7ex\hbox{E}\kern-.125emX}}

\begin{document}
\title{AI Generated Signal for Wireless Sensing
}
\author{Hanxiang~He\IEEEauthorrefmark{2},
Han~Hu\IEEEauthorrefmark{2},
Xintao~Huan\IEEEauthorrefmark{2},
Heng~Liu\IEEEauthorrefmark{2},
Jianping~An\IEEEauthorrefmark{2},
Shiwen~Mao\IEEEauthorrefmark{3}
\\
\IEEEauthorblockA{\IEEEauthorrefmark{2}Beijing Institute of Technology, China, {\{hxhe, hhu, xintao.huan, lhengzzt, an\}@bit.edu.cn}\\
\IEEEauthorrefmark{3}Auburn University, Auburn, AL, USA, {smao@ieee.org}
}

\thanks{Han Hu is the corresponding author.}

\vspace{-0.5cm}}

\maketitle

\begin{abstract}

Deep learning has significantly advanced wireless sensing technology by leveraging substantial amounts of high-quality training data. However, collecting wireless sensing data encounters diverse challenges, including unavoidable data noise, limited data scale due to significant collection overhead, and the necessity to reacquire data in new environments. Taking inspiration from the achievements of AI-generated content, this paper introduces a signal generation method that achieves data denoising, augmentation, and synthesis by disentangling distinct attributes within the signal, such as individual and environment. The approach encompasses two pivotal modules: structured signal selection and signal disentanglement generation. Structured signal selection establishes a minimal signal set with the target attributes for subsequent attribute disentanglement. Signal disentanglement generation disentangles the target attributes and reassembles them to generate novel signals. 
Extensive experimental results demonstrate that the proposed method can generate data that closely resembles real-world data on two wireless sensing datasets, exhibiting state-of-the-art performance.
Our approach presents a robust framework for comprehending and manipulating attribute-specific information in wireless sensing.

\end{abstract}


\section{Introduction}
Wireless sensing has emerged as a potent method for applications such as person identification, pose tracking, and gesture recognition due to its remarkable flexibility and robust privacy protection \cite{li2022,zhang2020}. Concurrently, deep learning serves as a formidable tool for identifying complex patterns and relationships within data \cite{ge2021, burgess2018, chen2018}. In the field of wireless sensing empowered by deep learning, numerous advantages are revealed compared to traditional methods, leading to improvements in accuracy, scalability, and energy efficiency. This establishes it as a highly promising technology with extensive potential for diverse applications \cite{piriyajitakonkij2020,li2022}.

The success of deep learning-based wireless sensing relies heavily on substantial volumes of high-quality data, which can present challenges due to various factors. First, wireless sensing data is susceptible to noise  and interference, potentially resulting in data loss or corruption. 
Second, deep learning algorithms demand significant amounts of high-quality data to facilitate learning and enhance their performance. The process of obtaining such data can be time-consuming, especially in scenarios involving human subjects, where ethical considerations and privacy concerns may arise. Third, the trained model is meticulously calibrated to data collected from a specific environment, which might not seamlessly transfer to different environments. Discrepancies in data distributions can introduce bias, ultimately impacting the accuracy and reliability of the models.
These challenges require careful consideration of data denoising, augmentation, and synthesis to improve the quality and diversity of collected data and transfer knowledge from seen to unseen scenarios for improving model performance. 1) Traditional signal denoising methods typically utilize linear and non-linear filtering to eliminate noise. However, these methods often struggle to differentiate between noise and signal components, potentially resulting in the removal of valuable information along with the noise. 2) Data augmentation methods employ various time-frequency transformations on existing data to generate novel samples while preserving similar data distribution. Li \textit{et al.} \cite{li2022} enriched time-domain characteristics by transforming signal from diverse perspectives, including distance, angle, speed, and trajectory. Zhang \textit{et al.} \cite{zhang2020} shifted signals from the time domain to the frequency domain and subsequently applied scaling and filtering techniques. Nonetheless, these techniques might result in the creation of unrealistic data, potentially leading to erroneous or biased training. 3) Data synthesis methods strive to generate signals with distinct distributions from the original dataset by leveraging the resemblances in physical laws between seen and unseen scenarios. Tang \textit{et al.} \cite{tang2021} utilized Variational Autoencoder (VAE) to transform the original samples into the latent space, capture the feature distribution, and generate target samples through distribution-based sampling. Ge \textit{et al.} \cite{ge2021} generated samples with arbitrary distributions by controlling the latent feature vector. However, these approaches require access to partial data from unseen environments to acquire prior knowledge, which proves infeasible in practical applications.

To address these limitations, we propose an innovative attribute disentanglement approach that tackles the challenges of data denoising, augmentation, and synthesis within wireless sensing scenarios. These scenarios typically involve multiple attributes, such as person, room, and barrier. The fundamental principle of our approach revolves around disentangling the signal features corresponding to these attributes, enabling us to learn compact and easily interpretable representations for each attribute. By leveraging these disentangled representations, we can effectively synthesize signals for diverse scenarios.

Our approach consists of two main modules: structured signal selection and signal disentanglement generation. In the structured data selection module, we carefully choose signal data groups that contain the necessary attribute features from an existing dataset. This process guarantees the inclusion of crucial information required for attribute disentanglement. Subsequently, in the data disentanglement generation module, we utilize an autoencoder neural network to extract disentangled attribute features and recombine them to generate desired signals. The autoencoder neural network can enable the separation of attribute-specific information from the input signals, facilitating the synthesis of artificial signals that reflect specific attribute combinations.

To the best of our knowledge, this work represents a pioneering effort in the controlled synthesis of signals using deep representation learning within the wireless sensing domain. 
By harnessing the capabilities of deep learning techniques, our approach not only effectively addresses the challenges of data denoising, augmentation, and synthesis but also establishes a robust framework for comprehending and manipulating attribute-specific information in wireless sensing scenarios.
Through the disentanglement of signal features and subsequent signal synthesis, our approach facilitates the development of resilient and adaptable applications within wireless sensing systems.
Additionally, we contribute to the research community by releasing the datasets and source codes associated with our work\footnote{\href{https://github.com/hxhebit/AIGSWS}{https://github.com/hxhebit/AIGSWS}}. 
\vspace{-0.3cm}

\section{Problem Formulation}

\begin{figure}[!tb]
    \setlength{\abovecaptionskip}{-0.1cm}
    \centerline{\includegraphics[width=.95\linewidth]{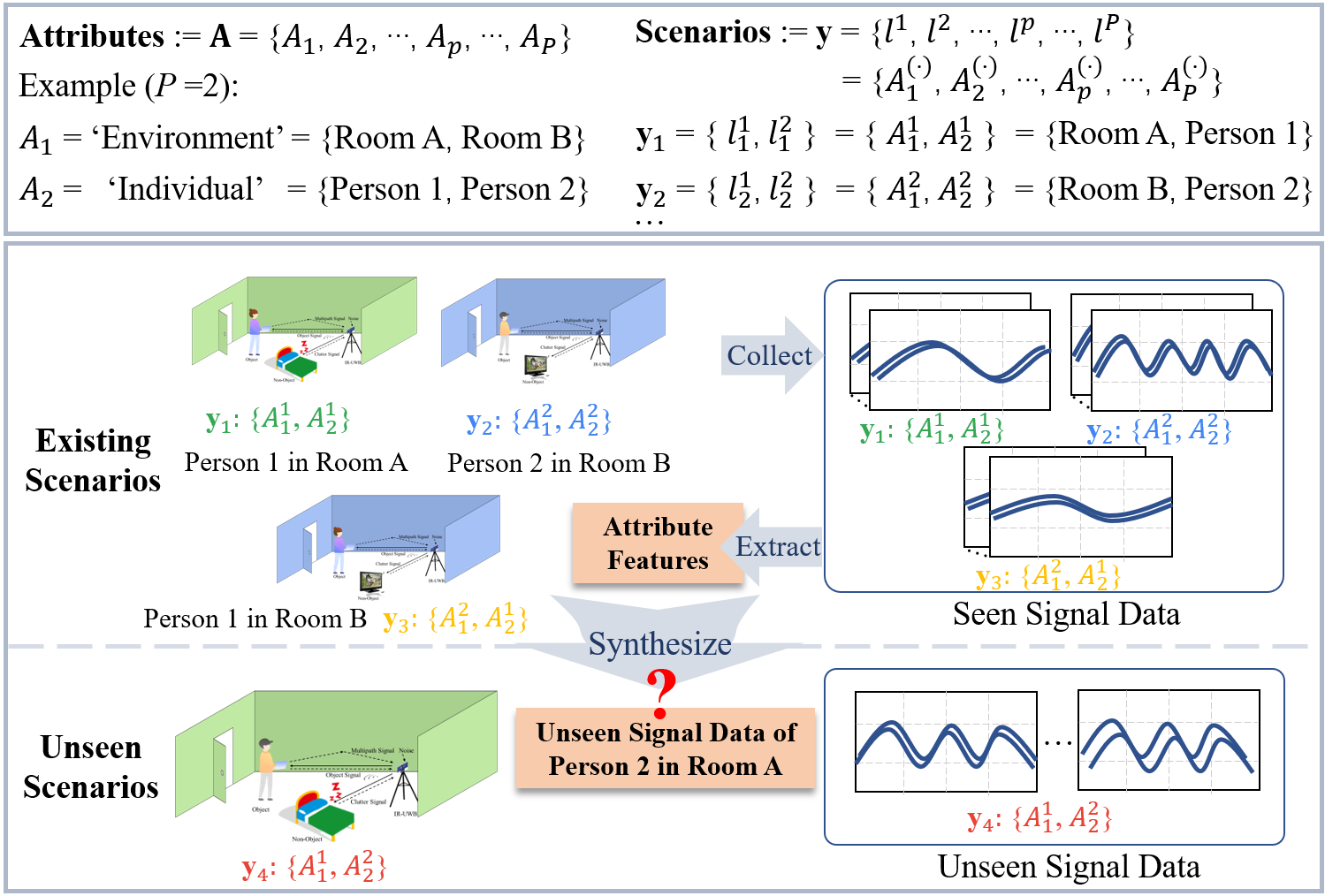}}
    \caption{A toy example of wireless sensing-based person identification.}  
    \label{fig:fig1}
    \vspace{-1.5em}
\end{figure}

\textbf{Preliminary Terms}. Fig.~\ref{fig:fig1} illustrates a toy example of person identification based on wireless sensing, introducing relevant terms and objective. The primary goal is to identify individuals within a room by leveraging wireless technologies like Wi-Fi or Radar. In this particular example, we consider two distinct \textbf{attributes}: `Environment' and `Individual'. The `Environment' attribute comprises two categories, i.e., Room A and Room B. Simultaneously, the `Individual' attribute includes two categories, namely Person 1 and Person 2. By combining these attribute categories, various \textbf{scenarios} emerge. For instance, the scenarios involve Person 1 in Room A, Person 1 in Room B, Person 2 in Room A and Person 2 in Room B, resulting in a total of four possible scenarios. It is essential to note that these scenarios can be further classified into two types: \textbf{existing scenarios} and \textbf{unseen scenarios}. Existing scenarios refer to situations for which data is already present in the collected dataset. In contrast, unseen scenarios encompass situations where the corresponding data has not been collected and needs to be generated.

\textbf{Mathematical Formulation}. The \text{attribute} set is denoted as $\mathbf{A} = \{A_1, A_2, \cdots, A_p, \cdots, A_{P}\}$, where the number of attributes is $P$, determined by the specific dataset. Each attribute $A_p$ has a category size denoted as $|A_{p}|$. A \text{scenario} $\mathbf{y} = \{l^{1},l^2,\ldots,l^p,\ldots,l^P\}$ represents a combination of $P$ attributes, where $l^{p} \in A_p$, indicating that $l^{p}$ belongs to the category sets associated with attribute $A_p$. The datasets of \text{existing scenarios} and \text{unseen scenarios} are represented as $\mathcal{B}$ and $\mathcal{V}$, respectively. The samples in each dataset are denoted as $\{(\mathbf{x}_i, \mathbf{y}_i)\}_{i=1}^{N_{Set}}$ , where $N_{Set}$ is the size of the dataset. The joint distribution of the dataset is represented by $P_{Set}$.

Regarding the task of signal synthesis, the objective is to construct a generator $G: \mathcal{B} \to \widetilde{\mathcal{V}}$ that is capable of generating unseen samples $(\widetilde{\mathbf{x}}, \widetilde{\mathbf{y}}) = G(\mathbf{x}^\mathcal{B}, \mathbf{y}^\mathcal{B}) \in \widetilde{\mathcal{V}}$, where $\widetilde{(\cdot)}$ signifies the synthesis of samples. The ultimate goal is to ensure similarity between the synthetic samples and the real samples, capturing the features and distribution of the original data. The final optimization target of signal synthesis is:
\vspace{-.5em}
\begin{equation}
     \min_{G} f(P_{\mathcal{V}}, P_{\widetilde{\mathcal{V}}}),
     \label{func:num}
     \vspace{-.5em}
 \end{equation}
where $f(\cdot)$ is the probability distribution distance measure function. Similarly, the objective of signal augmentation can be modeled as constructing a generator $G: \mathcal{B} \to \widetilde{\mathcal{B}}$ that $\min\limits_{G} f(P_{\mathcal{B}}, P_{\widetilde{\mathcal{B}}})$ to ensure the authenticity of generated signal.
\vspace{-0.5cm}

\section{Signal Generation Method}

\subsection{Overview}
Given an unseen scenario, we utilize attribute disentanglement to decompose it into distinct attributes. For instance, in Fig. \ref{fig:fig1}, for the unseen scenario $\mathbf{y}_4= \{A_1^1, A_2^2\}$, we construct separate training sets based on existing scenarios for attributes $A_1^1$ and $A_2^2$. Subsequently, we learn the independent representations of these attributes and then combine them to generate the desired signal. 

The proposed method comprises two modules: 
\begin{itemize}
    \item \textbf{Structured signal selection}. A set of samples is selected based on their attribute relationships, which is further categorized into three types as input for the subsequent neural network, facilitating the training of different components within the subsequent module.

    \item \textbf{Signal disentanglement generation}. An autoencoder neural network is designed and utilized for three components: feature extraction, feature disentanglement, and signal generation.
    Specifically, the type 1 sample from the selected set is encoded and then decoded to extract attribute features. Two type 2 samples are encoded, and their attribute features are exchanged for feature disentanglement. Finally, the disentangled features of two type 3  samples are combined to synthesize the desired signals.
\end{itemize}
\vspace{-0.3cm}

\subsection{Structured Signal Selection}
Although the unseen signals are unavailable, their attribute information can be acquired from the existing scenarios. Therefore, we can separately extract the attribute features and then combine them to synthesize the unseen signals. To extract the attribute features from the signals, we select paired signals that preserve the complete attribute information, enabling feature disentanglement and model optimization of subsequent neural network.

Since the unseen signal cannot be accessed during the training process, it cannot serve as a reference to evaluate the quality of the synthetic signal and optimize the model to generate high-quality samples. 
To address this, we introduce the Minimum Complete Information set $Set_{MCI}=\{\mathbf{x}_{m}, \mathbf{y}_{m}\}_{m=1}^{4}$ for feature disentanglement and model optimization, and for the samples within the set, the relationships hold: $l_{1}^{P-1} = l_{2}^{P-1}$, $l_{1}^{P} = l_{3}^{P}$, $l_{4}^{P-1} = l_{3}^{P-1}$, and $l_{4}^{P} = l_{2}^{P}$. Besides, the samples possess $P-2$ attributes of the same category when $P>2$. 
The samples within $Set_{MCI}$ are categorized into three types:
\begin{itemize}
    \item Type 1: Single sample, which represents each individual sample within the set.
    \item Type 2: Two samples with adjacent relationship, sharing one common attribute.
    \item Type 3: Two samples with diagonal relationship that do not share any common attribute.
\end{itemize}

It's worth noting that any newly generated signal, resulting from the exchange of attribute categories among the samples in this set, remains confined within the set. Therefore, we can utilize the samples within the set to evaluate the synthetic samples and optimize the model. Due to the structural characteristic, $Set_{MCI}$ is referred to as the structured signal.

\begin{figure}[!tb]
    \setlength{\abovecaptionskip}{0.cm}
    \centerline{\includegraphics[width=\linewidth]{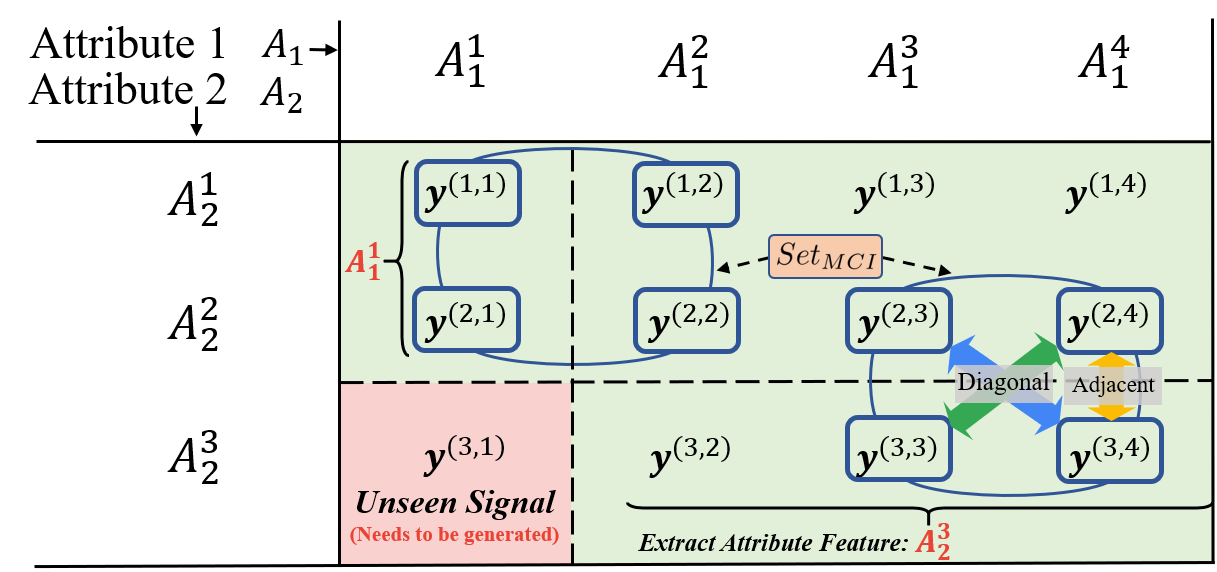}}
    \caption{An example of the structured signal selection method. The attribute categories of the sample are shown in the leftmost and topmost columns of the table. The category size $|A_{i}|$ is determined by the specific dataset, where the example shows $|A_{1}|=4$ and $|A_{2}|=3$.}
    \label{fig:fig4-2}
    \vspace{-1.5em}
\end{figure}

An example of the structured signal selection method is illustrated in Fig.~\ref{fig:fig4-2}.
For simplicity, the attribute size $P$ is set to 2, resulting in each sample $\mathbf{x}_i$ having two attributes, i.e., $A_1$ and $A_2$. The scenario $\mathbf{y}^{(p,q)}$ denotes the combination of attribute category, for instance, $\mathbf{y}^{(1,1)}$ represents the category combination of $\{A_1^1, A_2^1\}$. The unseen signal $\mathbf{y}^{(3,1)}$ is combined of $\{A^1_1, A^3_2\}$. The background color indicates the dataset to which the samples belong: red for the unseen dataset $\mathcal{V}$, and green for the existing dataset $\mathcal{B}$.
Two $Set_{MCI}$ are denoted by blue circles, from which we separately extract the attribute features of $A^1_1$ and $A^3_2$.

\vspace{-0.3cm}
\subsection{Signal Disentanglement Generation}
\vspace{-0.1cm}
We propose an autoencoder neural network for generating desired signal. As depicted in Fig.~\ref{fig:fig4-3}, three components of feature extraction, feature disentanglement, and signal synthesis utilize a shared autoencoder neural network for training.
Specifically, the network extracts features from input samples, further disentangles these features, and ultimately combines the separated attribute features to synthesize the target signals.

\subsubsection{Feature Extraction}
The network encodes the samples into a low-dimensional latent space, preserving the features necessary for generating the original samples and performing denoising. The inputs of this component are type 1 samples.

In the component 1 of Fig.~\ref{fig:fig4-3}, the autoencoder neural network $G_\mathbf{\theta_g}$ comprises an encoder $E_\mathbf{\theta_e}:\mathcal{X} \to \mathcal{Z} $ and a decoder $D_\mathbf{\theta_d}:\mathcal{Z}  \to \mathcal{X} $. The encoder $E_\mathbf{\theta_e}$ maps the input sample $\mathbf{x}_i$ to a latent feature space $\mathcal{Z} $, obtaining the feature vector $\mathbf{z}_i \in \mathbb{R} ^d$. Subsequently, the decoder generates reconstructed samples $\widetilde{\mathbf{x}}_i = D_\mathbf{\theta_d}(\mathbf{z}_i)$. To enhance the quality of generated sample $\widetilde{\mathbf{x}}_i$, the reconstruction loss is utilized to minimize the disparity between  $\widetilde{\mathbf{x}}_i$ and $\mathbf{x}_i$ as:
\vspace{-0.5em}
\vspace{-1.0em}

\begin{equation} 
    \vspace{-.5em}
    \mathcal{J}_{RECON} = \mathbb{E}_{\mathbf{x}_i\thicksim P_{\mathcal{B} }} || \mathbf{x}_i - D_\mathbf{\theta_d}(E_\mathbf{\theta_e}(\mathbf{x}_i)) ||_1.
    \label{func:recon}
\end{equation}

\vspace{-.5em}
\subsubsection{Feature Disentanglement}
The latent feature vector $\mathbf{z}$ can influence the category of attribute \cite{ge2021}. Consequently, we can synthesize samples with designated attribute categories by controlling the latent feature vector $\mathbf{z}$. To this end, we introduce the attribute feature exchange operation to exchange portions of the feature vectors between two samples, subsequently generating original samples. The inputs for this component are samples of type 2.

\begin{figure*}[]
    \setlength{\abovecaptionskip}{-0.2cm}
    \centerline{\includegraphics[width=.85\linewidth]{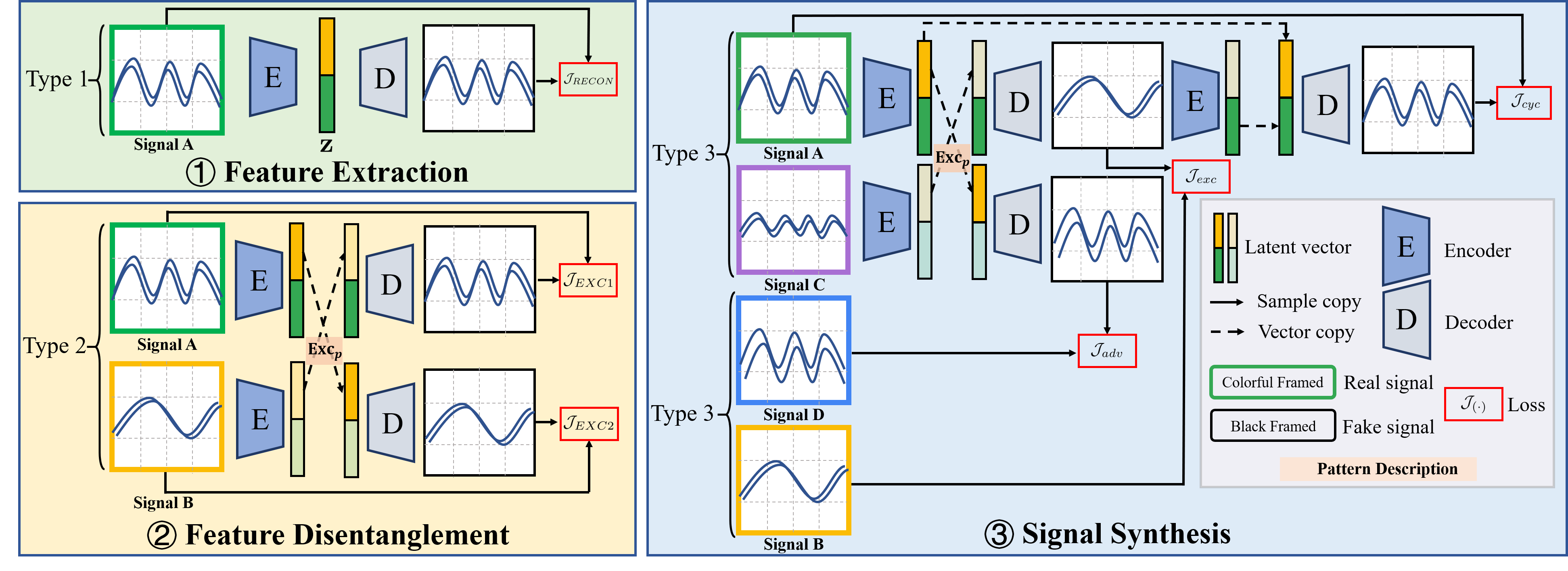}}
    \caption{The architecture of signal disentanglement generation. Three types of signal samples are input into the autoencoder neural network. And three distinct training components facilitate the neural network in acquiring the abilities of feature extraction, feature disentanglement, and signal synthesis, respectively.}
    \label{fig:fig4-3}
    \vspace{-1.5em}
\end{figure*}

The latent feature vector $\mathbf{z}$ is partitioned into $P$ segments, i.e., $\mathbf{z} =\{z^1, z^2,\dots,z^P\}$, where each sub-vector $z^p$ corresponds to an attribute $A_p$. This implies that the attribute category $l^p \in A_p$ is controlled by $z^p$. The attribute feature exchange operation $\mbox{Exc}_p$ is introduced to exchange sub-vector $z^p$ between samples $\mathbf{z}_i$ and $\mathbf{z}_j$ as follows:
\vspace{-0.8em}
\begin{equation}
    \vspace{-.8em}
    \begin{aligned}
        \mbox{Exc}_p(\mathbf{z}_i,\mathbf{z}_j) = [\mathbf{z}_{i}^{'}, \mathbf{z}_{j}^{'}] = \Big[\{z_i^1, z_i^2,\dots,z_j^p,\dots,z_i^P\}, \\
          \{z_j^1, z_j^2,\dots,z_i^p,\dots,z_j^P\}\Big].
    \end{aligned}
    \label{func:exc_op}
 \end{equation}

For feature disentanglement, as shown in component 2 of Fig.~\ref{fig:fig4-3}, the type 2 samples $\mathbf{x}_i$ and $\mathbf{x}_j$ (the yellow arrow in Fig.~\ref{fig:fig4-2}) are input into the encoder $E_\mathbf{\theta_e}$, getting the latent feature vectors $\mathbf{z}_i$ and $\mathbf{z}_j$.
Since samples $\mathbf{x}_i$ and sample $\mathbf{x}_j$ share the same attribute category for a particular attribute, exchanging the sub-vectors controlling this attribute will not alter the original attribute category combination, i.e., $\widetilde{\mathbf{y}}_i = \mathbf{y}_i$ and $\widetilde{\mathbf{y}}_j = \mathbf{y}_j$. This indicates the synthetic samples $\widetilde{\mathbf{x}}_i$ and $\widetilde{\mathbf{x}}_j$ can closely resemble the original samples as follows:
\vspace{-0.5em}
\begin{equation}
    \vspace{-0.5em}
    \widetilde{\mathbf{x}}_i = D_\mathbf{\theta_d}(\mathbf{z}_i^{'}) \thickapprox \mathbf{x}_i, \\
    \widetilde{\mathbf{x}}_j = D_\mathbf{\theta_d}(\mathbf{z}_j^{'}) \thickapprox \mathbf{x}_j,
    \label{func:exc_1}
\end{equation}
where $\mathbf{z}_i^{'} = Exc_k(\mathbf{z}_i,\mathbf{z}_j)[i]$, $\mathbf{z}_j^{'}= Exc_k(\mathbf{z}_i,\mathbf{z}_j)[j]$, and $k$ is the index of the attribute.

The reconstruction loss is utilized to assess the disparity between the synthesized original sample and the real original sample. The model can be optimized as follows:
\vspace{-0.5em}

\vspace{-1.0em}
\begin{equation}
    \begin{aligned}
    \vspace{-1.5em}
    & \mathcal{J}_{EXC1} = \mathbb{E}_{\mathbf{x}_i,\mathbf{x}_j\thicksim P_{\mathcal{B} }} || \mathbf{x}_i - D_\mathbf{\theta_d}(Exc_k(E_\mathbf{\theta_e}(\mathbf{x}_i,\mathbf{x}_j))[i]) ||_1,\\
    & \mathcal{J}_{EXC2} = \mathbb{E}_{\mathbf{x}_i,\mathbf{x}_j\thicksim P_{\mathcal{B} }} || \mathbf{x}_j - D_\mathbf{\theta_d}(Exc_k(E_\mathbf{\theta_e}(\mathbf{x}_i,\mathbf{x}_j))[j]) ||_1,\\
    & \mathcal{J}_{EXC} = \mathcal{J}_{EXC1} + \mathcal{J}_{EXC2}.
    \end{aligned}
    \label{func:exc}
\end{equation}
\subsubsection{Signal Synthesis}
We synthesize unseen data by exchanging attribute features between two samples with completely different attribute categories and further leverage additional accessible samples for model optimization. The inputs in this component are samples of type 3.

As depicted in component 3 of Fig.~\ref{fig:fig4-3}, the type 3 samples A and C (the blue arrow in Fig.~\ref{fig:fig4-2}) are input into the encoder $E_\mathbf{\theta_e}$ to obtain the latent feature vectors $\mathbf{z}$. After applying the operation $Exc_p(\mathbf{z}_i,\mathbf{z}_j)$, the unseen samples can be synthesized by the decoder $D_\mathbf{\theta_d}$. Note that, we cannot utilize the original input samples to evaluate the synthetic samples due to differences in attribute category combinations, i.e., $\widetilde{\mathbf{y}}_i\neq \mathbf{y}_i$ and $\widetilde{\mathbf{y}}_j\neq \mathbf{y}_j$. Therefore, the additional samples B and D are involved as the reference target (the green arrow in Fig.~\ref{fig:fig4-2}), where these samples have the same attributes as synthetic samples. Additionally, the synthetic sample undergoes encoding and attribute feature exchange to generate the original sample. Besides, the adversarial loss is introduce for model optimization. The final objective functions are formulated as:
\vspace{-2em}

\begin{equation}
    \begin{aligned}
        \vspace{-.5em}
       & \mathcal{J}_{exc} = \mathbb{E}_{\mathbf{x}_i,\mathbf{x}_t\thicksim P_{\mathcal{B}}} || \mathbf{x}_j - D_\mathbf{\theta_d}(Exc_p(E_\mathbf{\theta_e}(\mathbf{x}_i,\mathbf{x}_t))[j]) ||_1,\\
        &\mathcal{J}_{cyc} = \mathbb{E}_{\mathbf{x}_i,\mathbf{x}_t\thicksim P_{\mathcal{B} }} || \mathbf{x}_i - D_\mathbf{\theta_d}(Exc_p(E_\mathbf{\theta_e}(\widetilde{\mathbf{x}}_j),\mathbf{z}_i)[i] )||_1,\\
        &\mathcal{J}_{adv}  = \mathbb{E}_{\widetilde{\mathbf{x}}_{l}\thicksim P_{\widetilde{\mathcal{B} }}}[log Q_\mathbf{\theta_q}(\widetilde{\mathbf{x}}_{l})] + \mathbb{E}_{\mathbf{x}_{l}\thicksim P_{\mathcal{B} }}[log(1 - Q_\mathbf{\theta_q}(\mathbf{x}_{l}))],\\
        & \mathcal{J}_{GEN} = \mathcal{J}_{exc} + \gamma \mathcal{J}_{cyc} + \lambda \mathcal{J}_{adv},
    \end{aligned}
    \label{func:gen}
\end{equation}
where $ \gamma$ and $\lambda$ are hyper-parameters, and $Q_{\mathbf{\theta_q}}(\cdot)$ denotes a neural network designed to differentiate between real and synthetic samples.

\subsubsection{Training Process}
The training process is as follows:
\begin{itemize}
    \item In the feature extraction component, each sample of type 1 is encoded into a low-dimensional latent space and then decoded to calculate $\mathcal{J}_{RECON}$.
    \item In the feature disentanglement component, samples of type 2 (the yellow arrow in Fig.~\ref{fig:fig4-2}) are separately encoded to obtain their feature vectors, which are partially swapped and recombined to generate the original samples for calculating the $\mathcal{J}_{EXC}$.
    \item In the signal synthesis component, samples of type 3 (the blue/green arrows in Fig.~\ref{fig:fig4-2}) are encoded and their latent vectors are swapped to generate another diagonal sample (the green/blue arrows in Fig.~\ref{fig:fig4-2}), which are compared with the real samples to calculate $\mathcal{J}_{GEN}$.
\end{itemize}
Finally, the optimization object is:
\begin{equation}
    \vspace{-.5em}
    \mathcal{J}_{ALL} = \mathcal{J}_{RECON} + \alpha \mathcal{J}_{EXC} + \beta \mathcal{J}_{GEN},
    \label{func:all}
\end{equation}
where $\alpha > 0$ and $\beta > 0$ are hyper-parameters for balancing the optimization function.
$\mathcal{J}_{RECON}$ promotes the encoder to extract effective features, $\mathcal{J}_{EXC}$ facilitates the encoder to extract disentangled features, and $\mathcal{J}_{GEN}$ enhances decoder's generation ability.
\vspace{-0.3cm}

\section{Experiments}
\begin{table*}[]
    \setlength{\abovecaptionskip}{-0.1cm} 
    \centering
    \label{tab:tab_1}
    \caption{Classification Result on the Synthetic Signal Dataset}
    \setlength{\tabcolsep}{2.2mm} 
    \renewcommand{\arraystretch}{1.2} 
    {
    \begin{tabular}{cc|cccc|cccc}
    \bottomrule[0.8pt]
    \multicolumn{1}{l}{}                                                                        & \multicolumn{1}{c|}{Task}                          & \multicolumn{4}{c|}{Classification for   Attribute 1 (Environment)} & \multicolumn{4}{c}{Classification for   Attribute 2 (Individual)} \\
    \multicolumn{1}{l}{}                                                                        & \multicolumn{1}{c|}{Method}                        & Accuracy(\%)    & Precision(\%)  & Recall(\%)     & \multicolumn{1}{c|}{F1(\%)}         & Accuracy(\%)    & Precision(\%)  & Recall(\%)     & F1(\%)         \\ \hline
    \multirow{3}{*}{\begin{tabular}[c]{@{}c@{}}with   \\      priori knowledge\end{tabular}}    & \multicolumn{1}{c|}{Original}                      & 77.22           & 77.24          & 77.22          & \multicolumn{1}{c|}{77.40}          & 81.67          & 81.57          & 81.67          & 81.97          \\ 
                                                                                                & \multicolumn{1}{c|}{SR-RFFI\cite{shen2022}}        & 73.33           & 73.36          & 73.33          & \multicolumn{1}{c|}{74.78}          & 79.44          & 79.08          & 79.44          & 79.68          \\
                                                                                                & \multicolumn{1}{c|}{SPN\cite{piriyajitakonkij2020}}        & 73.89           & 73.38          & 73.89          & \multicolumn{1}{c|}{74.33}          & 79.44          & 79.17          & 79.44          & 79.62          \\ \hline
    \multirow{4}{*}{\begin{tabular}[c]{@{}c@{}}without   \\      priori knowledge\end{tabular}} & \multicolumn{1}{c|}{$\beta$-VAE\cite{burgess2018}} & 70.56           & 70.30          & 70.56          & \multicolumn{1}{c|}{71.94}          & 78.89          & 78.76          & 78.89          & 79.59          \\
                                                                                                & \multicolumn{1}{c|}{btcVAE\cite{chen2018}}         & 69.44           & 68.93          & 69.44          & \multicolumn{1}{c|}{71.62}          & 78.89          & 78.57          & 78.89          & 79.64          \\
                                                                                                & \multicolumn{1}{c|}{GZS-Net\cite{ge2021}}          & 68.33           & 68.06          & 68.33          & \multicolumn{1}{c|}{69.31}          & 78.89          & 78.56          & 78.89          & 78.72          \\
                                                                                                & \multicolumn{1}{c|}{Ours}                          & \textbf{73.33}  & \textbf{73.35} & \textbf{73.33} & \multicolumn{1}{c|}{\textbf{74.27}} & \textbf{80.00} & \textbf{79.67} & \textbf{80.00} & \textbf{80.84} \\
                                                                             \bottomrule[0.8pt] 
    \end{tabular}}
    \vspace{-0.5cm}
\end{table*}

\subsection{Experimental Setup}
\subsubsection*{\textbf{Datasets}}
A self-collected dataset and a public dataset are used to evaluate the model. 
The Person-UWB dataset is constructed for person identification and was gathered using Impulse Radio Ultra-WideBand Radar (Xethru X4M03). It comprises 900 samples from 5 individuals across 3 different environments, with attributes labeled as `Environment' and `Individual'. The Dop-Net \cite{ritchie2020} dataset is dedicated to gesture recognition, including 4 gesture types performed by 6 individuals, resulting in a dataset of 2433 samples. The attributes of this dataset are `Individual' and `Gesture'.
The datasets are partitioned into training and testing sets using an ratio of 8:2. Our approach is evaluated on both datasets; however, due to space limitations, each experimental section presents results from a single dataset.

\subsubsection*{\textbf{Baselines}}
We select five representative methods for fair comparisons of data augmentation and data generation, some of which require access to unseen scenarios as extra prior knowledge, which mitigates the difficulty in data generation.
\begin{itemize}
    \item SR-RFFI\cite{shen2022}: A wireless signal data augmentation method based on enriching channel diversity.
    \item SPN\cite{piriyajitakonkij2020}: A radar signal data augmentation method, shifting the signal on the temporal dimension and spatial dimension.
    \item $\beta$-VAE\cite{burgess2018}: A data generation method utilizing hyperparameters to enhance disentanglement capability.
    \item btcVAE\cite{chen2018}: A data generation method that balances the model's disentanglement and reconstruction abilities.
    \item GZS-Net\cite{ge2021}: A data generation method based on paired reconstruction.
\end{itemize}
\subsubsection*{\textbf{Parameters}}
\vspace{0.cm}
The encoder $E_\mathbf{\theta_e}$ consists of 4 concatenated modules of Convolution, Instance Normalization, and ReLU, followed by a residual module. The output is then mapped to 100 dimensions using 3 fully connected layers. The decoder $D_\mathbf{\theta_d}$ mirrors the architecture of the encoder $E_\mathbf{\theta_e}$. To ensure a fair comparison, identical architectures are used across all comparative methods. Adam optimizer with momentums of 0.9 and 0.999 is utilized, and the learning rate is set to 1e-4. Each sample is reshaped to $128\times128$ and every batch contains 3 samples. The model is trained for 150k iterations. The hyper-parameter $\alpha$, $\beta$, $\gamma$ and $\lambda$ are 1, 1, 0.2 and 0.1, respectively. 

\vspace{-0.3cm}
\subsection{Verifying Effectiveness of Feature Disentanglement}

We evaluate the disentanglement performance of our proposed model by estimating the correlation between latent features and their corresponding attributes. Specifically, we select two signals sharing the same attribute $A_k$, where one serves as the source signal and the other as the reference signal. Both signals are input into the trained model, producing their respective latent vectors $\mathbf{z}$. We then substitute the sub-vector $z_i^k$, which governs attribute $A_k$ in the source signal, with the corresponding segment from the latent sub-vector $z_j^k$ of the reference signal. Because the attribute categories governed by sub-vectors $z_i^k$ and $z_j^k$ are the same, exchanging them can retain the original attribute category combinations. When effective disentanglement is realized, the generated signal should perfectly match the original source signal.

\begin{figure}[!bp]
    \vspace{-1.5em}
    \setlength{\abovecaptionskip}{0.cm}
    \centerline{\includegraphics[width=.7\linewidth]{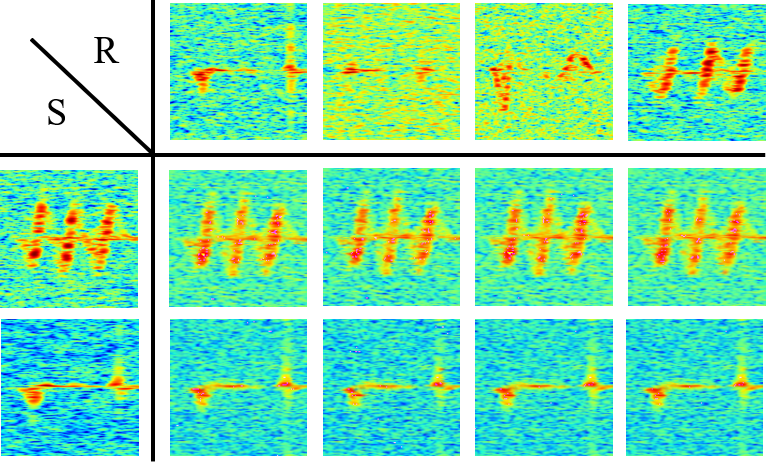}}
    \caption{Results of disentanglement performance evaluation. The source samples (S) are depicted on the left, while the reference samples (R) are on the top. The synthesized source samples are presented in the center.}
    \label{fig:fig5-1}
\end{figure}

The results on the Dop-Net\cite{ritchie2020} dataset are illustrated in Fig.~\ref{fig:fig5-1}, where the centrally positioned synthesized source signals closely resemble the actual source signals on the left. 
Especially, the doppler frequency shift induced by the gesture action remains unchanged in the synthesized source samples.
The similarity validates the efficacy of disentanglement.

\vspace{-0.5cm}
\subsection{Synthesizing Signal for Unseen Scenario}
Based on the feature disentanglement, we can control the latent feature vectors to synthesize signals of unseen scenarios. We first synthesize unseen signals and then utilize them to train an attribute classification model for quality evaluation.

\subsubsection{Signal Synthesis}

The signals of Person 1 in Corridor are synthesized, which is the unseen scenario during the training phase. 
Specifically, we select the reference signals that contain the attribute features of Person 1 or Corridor from the existing dataset. Then we severally extract these latent feature vectors and further combine them to synthesize the signals of Person 1 in Corridor.
\begin{figure*}[!tb]
    \setlength{\abovecaptionskip}{-.1cm}
    \centering
    \subfigure[]{
        \label{fig:fig5-2_1}
        \includegraphics[width=0.48\linewidth]{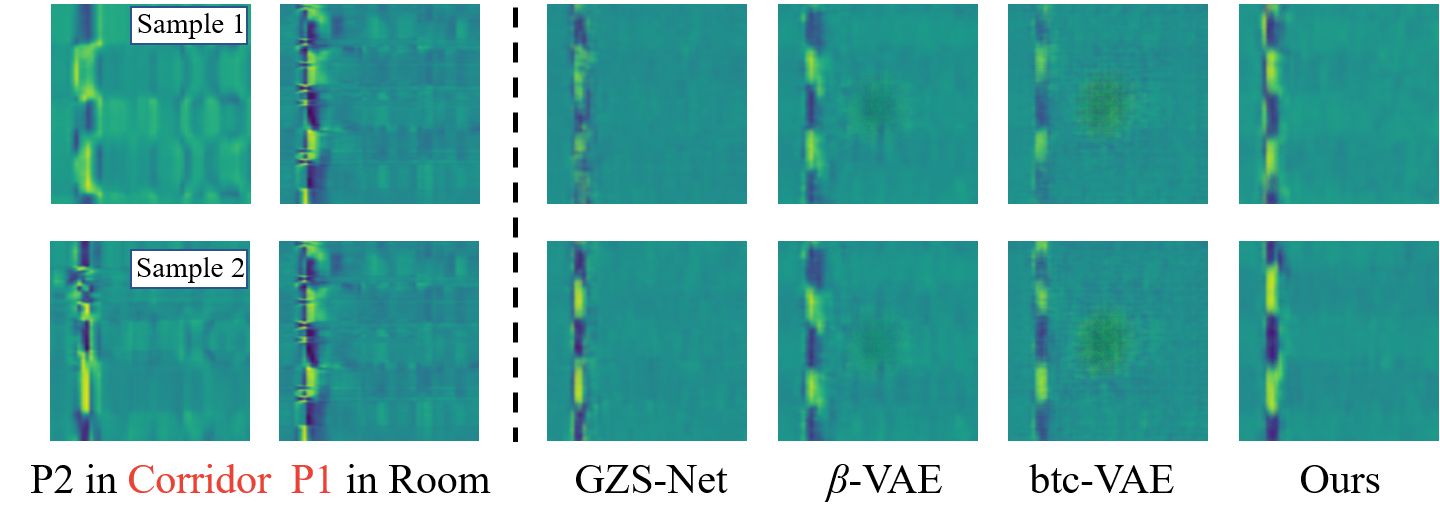}
        \vspace{-0.6cm}
    }
    \subfigure[]{
        \label{fig:fig5-2_2}
        \includegraphics[width=0.48\linewidth]{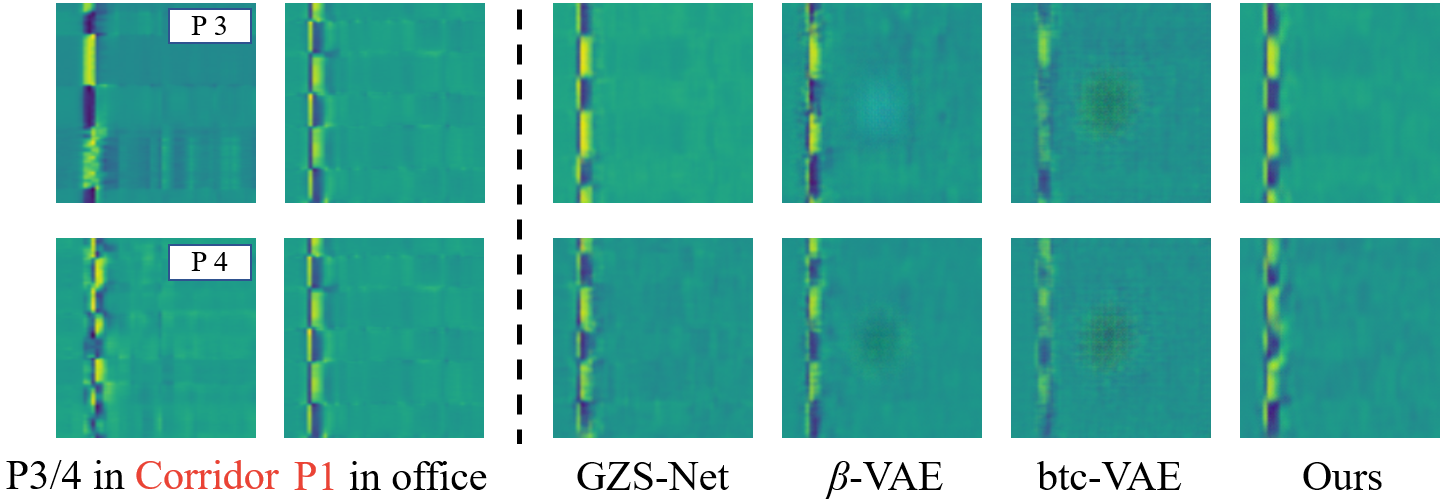}
        \vspace{-0.6cm}
    }
    \caption{Synthetic signal results of Person 1 in Corridor, where Person($\cdot$) is abbreviated to P($\cdot$). (a) Synthetic results that based on person 2 in the corridor and person 1 in the room. (b) Synthetic results that based on person 3/4 in the corridor and person 1 in the office. In both sub-figures, the first column represents the reference real signal of the `Environment' attribute (Corridor), the second column represents the reference real signal of the `Individual' attribute (P1), and the remaining columns represent the synthetic results.}
    \label{fig:fig5-2}
    \vspace{-1.2em}
 \end{figure*}

The synthesis results for Person 1 in the corridor are depicted in Fig.~\ref{fig:fig5-2}. Obviously, our method surpasses the baselines in terms of structural fidelity and clarity. As the slight movements like respiration and heartbeat of individuals have a more pronounced impact on signals in comparison to the static environment, the synthetic signals show a higher resemblance to the reference signals with individual attributes. Furthermore, despite the presence of noise in actual signals, our method is capable of generating clear signals.

\subsubsection{Signal Evaluation}

To evaluate the similarity between the synthetic unseen signals and the real unseen signals, we perform classification tasks separately for environment attribute and individual attribute. The results of the model trained on the synthetic signals are compared with those of the model trained on the real original signals. First, we construct the training dataset by combining the synthetic unseen signals with the existing signals, resulting in a complete training dataset. Then, we train a 3-layer fully connected network on the training dataset and test it on the real test dataset. The metrics of Accuracy, F1-score, Precision, and Recall are utilized for the evaluation. It is important to note that some baselines acquire access to the unseen signals as extra prior knowledge during the training process, which can mitigate the difficulty in data generation.

As shown in Table~I, it is evident that the results obtained by our method exhibit the highest proximity to the metrics of the original dataset. The accuracy shows only a slight decrease of 3.89\% and 1.67\% for the two classification tasks when compared to the original dataset. Remarkably, our method even outperforms the methods that incorporate prior knowledge in certain metrics. This indicates that our synthetic training set exhibits a data distribution comparable to that of the real training set.
\vspace{-0.45cm}
\subsection{Synthesizing Signal for Augmenting Seen Scenario}
For the signals of existing scenarios that we have collected in advance, our approach can further synthesize additional signals with identical attributes to augment the original dataset.

\subsubsection{Signal Synthesis}
\begin{figure}[!bp]
    \vspace{-1.5em}
    \setlength{\abovecaptionskip}{-.2cm}
    \centerline{\includegraphics[width=0.8\linewidth]{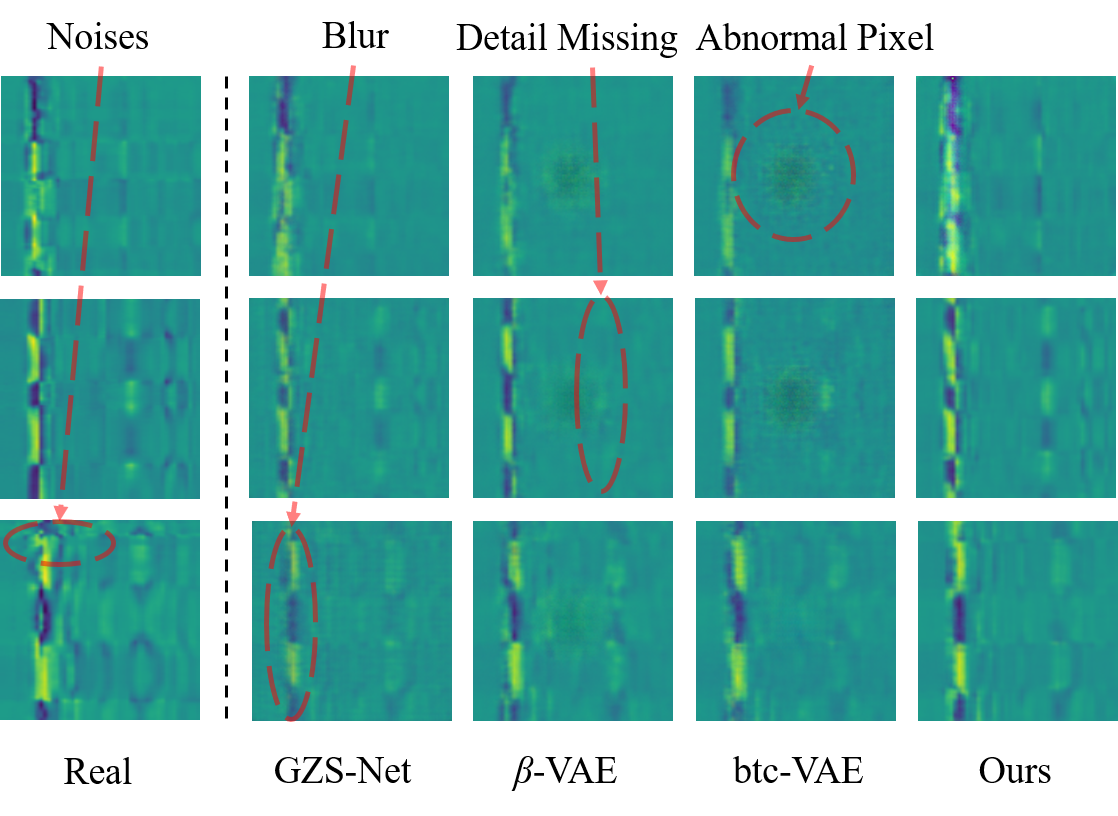}}
    \caption{Synthetic signal of the existing dataset. The first column depicts the real samples, and all other columns on the right side show the synthetic results.}
    \label{fig:fig5-3}
\end{figure}

The visual results on the Person-UWB dataset are depicted in Fig.~\ref{fig:fig5-3}. The signals synthesized by GZS-Net appear less sharp and more blurry, while $\beta$-VAE often misses the details present in real signals. Besides, btc-VAE tends to synthesize signals with numerous abnormal pixels. In comparison, our approach achieves superior results in terms of sharpness, contrast, and brightness. Importantly, our method effectively mitigates the impact of noise in real samples.

\subsubsection{Signal Evaluation}
We further evaluate the quantitative disparities between the synthetic signals and real signals. The similarity measurement metrics, Average Peak Signal-to-Noise Ratio (PSNR) and Structural Similarity Index (SSIM), are introduced to evaluate the synthesis performance ($\uparrow$ better). As presented in Table~II, our method achieves the best results in both two metrics, with values of 27.0100 and 0.8753, respectively. This confirms that the samples synthesized by our method closely resemble the real signals.
\vspace{-0.3cm}

\begin{table}[!tb]
    \setlength{\abovecaptionskip}{-0.1cm}
    \vspace{-1em}
    \centering
    \label{tab:tab_2}
    \caption{Comparison Results of the Real and Synthetic Signals}
    \setlength{\tabcolsep}{7mm}
    \renewcommand{\arraystretch}{1.2}{
    \begin{tabular}{ccc}
        \toprule[0.8pt]
            Method    & PSNR $\uparrow$    & SSIM $\uparrow$  \\ \hline
        \multicolumn{1}{c|}{$\beta$-VAE\cite{burgess2018}}    & 25.1645  & 0.8429               \\
        \multicolumn{1}{c|}{btcVAE\cite{chen2018}}    & 25.2259 & 0.8358               \\
        \multicolumn{1}{c|}{GZS-Net\cite{ge2021}}     & 25.0400 & 0.8364            \\ \hline
        \multicolumn{1}{c|}{Ours}   & \textbf{27.0100} & \textbf{0.8753}             \\
    \bottomrule[0.8pt]
    \end{tabular}}
    \vspace{-1em}
    \vspace{-0.3cm}
\end{table}

\section{Conclusion}
In this paper, we propose an signal generation framework for wireless sensing. The framework consists of structured signal selection and signal disentanglement generation components. Precisely, attribute features of unseen signals are extracted from the existing dataset and recombined to synthesize the desired signals. The method delivers substantial benefits in data denoising, augmentation, and synthesis. Importantly, our generation model operates independently of target data access during training.  In our forthcoming endeavors, we intend to broaden the scope of our approach to encompass intricate scenarios involving a greater variety of categories and attributes.

\vspace{.2cm}
\section*{Acknowledgement}
This work was supported by National Key R\&D Program of China under Grant 2022YFC3301201.
\vspace{.2cm}

\bibliographystyle{IEEEtran}
\normalem
\bibliography{IEEEabrv,ref}

\begin{thebibliography}{1}
\providecommand{\url}[1]{#1}
\csname url@samestyle\endcsname
\providecommand{\newblock}{\relax}
\providecommand{\bibinfo}[2]{#2}
\providecommand{\BIBentrySTDinterwordspacing}{\spaceskip=0pt\relax}
\providecommand{\BIBentryALTinterwordstretchfactor}{4}
\providecommand{\BIBentryALTinterwordspacing}{\spaceskip=\fontdimen2\font plus
\BIBentryALTinterwordstretchfactor\fontdimen3\font minus
  \fontdimen4\font\relax}
\providecommand{\BIBforeignlanguage}[2]{{%
\expandafter\ifx\csname l@#1\endcsname\relax
\typeout{** WARNING: IEEEtran.bst: No hyphenation pattern has been}%
\typeout{** loaded for the language `#1'. Using the pattern for}%
\typeout{** the default language instead.}%
\else
\language=\csname l@#1\endcsname
\fi
#2}}
\providecommand{\BIBdecl}{\relax}
\BIBdecl

\bibitem{li2022}
Y.~Li, D.~Zhang, J.~Chen, J.~Wan, D.~Zhang, Y.~Hu, Q.~Sun, and Y.~Chen,
  ``Di-gesture: Domain-independent and real-time gesture recognition with
  millimeter-wave signals,'' in \emph{Proc. IEEE GLOBECOM 2022, Rio, de
  Janeiro. Brazil, Dec.}, 2022, pp. 5007--5012.

\bibitem{zhang2020}
J.~Zhang, F.~Wu, B.~Wei, Q.~Zhang, H.~Huang, S.~W. Shah, and J.~Cheng, ``Data
  augmentation and dense-lstm for human activity recognition using wifi
  signal,'' \emph{IEEE Internet of Things Journal}, vol.~8, no.~6, pp.
  4628--4641, Mar. 2020.

\bibitem{ge2021}
Y.~Ge, S.~Abu-El-Haija, G.~Xin, and L.~Itti, ``Zero-shot synthesis with
  group-supervised learning,'' in \emph{International Conference on Learning
  Representations}, 2021.

\bibitem{burgess2018}
C.~P. Burgess, I.~Higgins, A.~Pal, L.~Matthey, N.~Watters, G.~Desjardins, and
  A.~Lerchner, ``Understanding disentangling in $\beta$-vae,'' \emph{arXiv
  preprint arXiv:1804.03599}, 2018.

\bibitem{chen2018}
R.~T. Chen, X.~Li, R.~B. Grosse, and D.~K. Duvenaud, ``Isolating sources of
  disentanglement in variational autoencoders,'' \emph{Advances in neural
  information processing systems}, vol.~31, 2018.

\bibitem{piriyajitakonkij2020}
M.~Piriyajitakonkij, P.~Warin, P.~Lakhan, P.~Leelaarporn, N.~Kumchaiseemak,
  S.~Suwajanakorn, T.~Pianpanit, N.~Niparnan, S.~C. Mukhopadhyay, and
  T.~Wilaiprasitporn, ``Sleepposenet: multi-view learning for sleep postural
  transition recognition using uwb,'' \emph{IEEE Journal of Biomedical and
  Health Informatics}, vol.~25, no.~4, pp. 1305--1314, 2020.

\bibitem{tang2021}
C.~Tang, W.~Li, S.~Vishwakarma, F.~Shi, S.~J. Julier, and K.~Chetty, ``Fmnet:
  Latent feature-wise mapping network for cleaning up noisy micro-doppler
  spectrogram,'' \emph{IEEE Transactions on Geoscience and Remote Sensing},
  vol.~60, pp. 1--12, Oct. 2021.

\bibitem{shen2022}
G.~Shen, J.~Zhang, A.~Marshall, and J.~R. Cavallaro, ``Towards scalable and
  channel-robust radio frequency fingerprint identification for lora,''
  \emph{IEEE Transactions on Information Forensics and Security}, vol.~17, pp.
  774--787, 2022.

\bibitem{ritchie2020}
M.~Ritchie, R.~Capraru, and F.~Fioranelli, ``Dop-net: a micro-doppler radar
  data challenge,'' \emph{Electronics Letters}, vol.~56, no.~11, pp. 568--570,
  2020.

\end{thebibliography}

\end{document}